\documentclass{aa}
\usepackage{graphics}
\usepackage{times}
\begin{document}
\thesaurus{08(09.04.1;09.08.1;09.09.1;13.09.3)}
%

%%%%%%%%%%%%%%%%%%%%%%%% Title Page begins %%%%%%%%%%%%%%%%%%%%%%%%%%%%%

\title{Far infrared observations of the southern Galactic star forming
complex around IRAS 09002-4732}

\author{S.K. Ghosh\inst{1} \and B. Mookerjea\inst{1} \and 
T.N. Rengarajan\inst{1} \and S.N. Tandon\inst{2} \and R.P. Verma\inst{1} }

\institute{Tata Institute of Fundamental Research, Homi Bhabha Road,  
Mumbai (Bombay) 400 005, India \and Inter-University Centre for Astronomy 
\& Astrophysics, Ganeshkhind, Pune 411 007, India}

\date{Received 9 May 2000 / Accepted 2 October 2000}
\offprints{S.K. Ghosh}
\mail{swarna@tifr.res.in}
\titlerunning{FIR observations of the IRAS 09002-4732 region}
\authorrunning{S.K. Ghosh et al.}
\maketitle
%%%%%%%%%%%%%%%%%%%%%%%%%%% Title Page ends %%%%%%%%%%%%%%%%%%%%%%%%%%%%%

%%%%%%%%%%%%%%%%%%%%%%%%%% Abstract begins %%%%%%%%%%%%%%%%%%%%%%%%%%%%%

\begin{abstract}

  The Galactic star forming region in the southern sky,
associated with IRAS 09002-4732 
has been mapped simultaneously in two
far infrared bands ($\lambda_{eff}$ = 148 \& 209 $\mu$m),
with $\sim$ 1\arcmin ~angular resolution. 
Fifteen sources, including 
IRAS 08583-4719, 08589-4714, 09002-4732 and 09014-4736 
have been detected, some of which are well resolved.
Taking advantage of similar beams in the two bands,
a reliable dust temperature [T(148/209)] map has been obtained, which
detects colder dust ($<$ 30 K) in this region.
The HIRES processed IRAS maps
at 12, 25, 60 and 100 $\mu$m, have also been used for comparison.
The optical depth maps, $\tau_{200}$ and $\tau_{100}$, 
generated from these far-IR data quantify the spatial distribution of the dust.
The diffuse emission from this entire region has been found to be 35 \% 
of the total FIR luminosity.
The slope of the IMF in the mass range 4--16 $M_{\odot}$ has
been estimated to be $-1.25^{+0.75}_{-0.65}$ 
for this star forming complex.

Radiative transfer models in spherical geometry have been explored to fit
available observations of the 4 IRAS sources 
and extract various 
physical parameters for corresponding dust-gas clouds.
Whereas a constant ($r^{0}$) radial density distribution is favoured in 
IRAS 08583-4719, 08589-4714 and 09002-4732,
the $r^{-1}$ law is inferred for IRAS 09014-4736.
The dust composition is found to be similar (Silicate dominated) 
in all the four IRAS sources modelled.
The luminosity per unit mass is found to be in the narrow range
of  44 -- 81 $L_{\odot}/M_{\odot}$  for these star forming regions.


\end{abstract}

%%%%%%%%%%%%%%%%%%%%%%%%%% Abstract ends %%%%%%%%%%%%%%%%%%%%%%%%%%%%%%%

%%%%%%%%%%%%%%%%%%%%%%%% Main Text begins %%%%%%%%%%%%%%%%%%%%%%%%%%%%%%

\section{Introduction}

   The study of far infrared (FIR) emission from the relatively denser regions
of the Galactic interstellar medium (ISM) is useful for understanding the
star formation process. High spatial resolution mapping of Galactic 
molecular clouds in the FIR continuum (in $\ge$ 2 bands)
provides direct information about the physical conditions 
(e.g. total luminosity, dust distribution, dust temperature etc.) 
of the region.
For a constant gas-to-dust
ratio, the dust optical depth traces matter along the line of sight. 
A comparison of
the dust temperature distribution with that of the gas temperature
(obtained from molecular line measurements) can give important information
on the dust-gas coupling in star forming regions. For medium to high
mass embedded YSOs, dust is the dominant source of heating of the gas
(outside the \ion{H}{ii} region).
A FIR study of an individual star formation complex (at a known distance)
can measure the luminosity function above a threshold which can usually
be translated to the present day mass function and even initial mass 
function.

Systematic differences can be expected
between star forming regions at different galactocentric distances,
since the star formation process depends on the
physical conditions of the local ISM (e.g. density, composition, radiation 
field etc.). 
This important issue has been addressed by many authors using various 
techniques. Systematic studies in radio continuum and molecular line
emission based on colour selected samples from 
the IRAS Point Source Catalog (hereafter IRAS PSC)
have led to very useful databases
which can be the starting point for further studies 
(Wood \& Churchwell \cite{WC89a}; Wouterloot \& Brand \cite{Wo89}).

 Although the IRAS mission has proved to be a corner stone in improving 
our understanding of star formation in the Galaxy, its sub-optimal angular 
resolution in the 60 and 100 $\mu$m bands has been a major limitation. 
In addition, recent ISO measurements 
of nearby spiral galaxies have shown the importance 
of emission at wavelengths $>$ 100 $\mu$m and colder dust (T $<$ 25 K) seems
to be a dominant constituent of spiral galaxies in general
(Alton et al. \cite{Al98}).
This suggests a similar role for colder dust in our Galaxy also.

In view of the above,
the TIFR 1-metre balloon borne telescope is flown regularly to 
map Galactic star forming regions simultaneously in two FIR bands
with near diffraction limited angular resolution.
These FIR bands have been chosen to be trans-IRAS
($\lambda_{eff}$ = 148~ \& ~209 $\mu$m), so that the
distribution of interstellar dust cooler than $\approx$ 30 K
which might have been missed by IRAS, can be traced
(Ghosh \cite{Gh00}, Mookerjea et al. \cite{Mo99}, \cite{Mo00a}, \cite{Mo00b}).
Accessibility of more than 90 \% of the Galactic plane 
($l$ = 315${}^{0}$ -- 0${}^{0}$ -- 290${}^{0}$)
from the Indian balloon launching station
has been exploited by targeting the 
southern sky, which is observationally
less explored in the FIR. 

 The catalogue of CO emission associated with colour selected
IRAS PSC sources by
Wouterloot \& Brand (\cite{Wo89}; hereafter WB89) covers the 
longitude range of the Galactic plane of interest to us. 
 Based on information from this catalogue,
the star forming complex around IRAS 09002-4732
was selected for the present study because : 
(i) all the four bright IRAS PSC sources in this region have similar
kinematic distances (WB89; Sugitani \& Ogura \cite{Su94}); three of them
(IRAS 08589-4714, 09002-4732 and 09014-4736)
are known to be physically associated, hence
this complex presents an opportunity to derive its PDMF / IMF if
a reasonable number of sources are detected;
(ii) it lies marginally
outside the solar circle, hence a comparison of the physical conditions 
in this region with that in the 4--5 kpc molecular region and regions
inside the solar circle would be interesting; 
(iii) the complex is very close to the  Galactic mid plane 
with $z \approx 20$ pc (WB89) as compared to the scale height
of 90 pc for the embedded OB stars (Wood \& Churchwell \cite{WC89b}).
This region around IRAS 09002-4732 is associated with 
the very extended Vela Molecular Ridge 
detected in the wide angle survey of $^{12}$CO
emission by May et al. (\cite{May88}) and Murphy and May (\cite{Mur91}),
and more recently mapped by Yamaguchi et al. (\cite{Ya99})
in both $^{12}$CO and $^{13}$CO lines.

Most of the available observational data for this region 
pertain to
IRAS 09002-4732 itself and clearly indicate the presence of an embedded
young stellar object in a dense molecular cloud.
These include measurements of
radio continuum (Caswell \& Haynes \cite{Ca87};
Manchester \& Goss \cite{Ma69}; Whiteoak \cite{Wh92};
Walsh et al. \cite{Wa98}),
infrared spectra from the IRAS Low Resolution Spectrometer 
Catalog (hereafter LRS; Olnon \& Raimond, \cite{Ol86}), ISO 
Long Wavelength Spectrometer data (ISO-LWS; 
Clegg et al. \cite{Cl96}) and photometric and imaging 
measurements in the near and mid infrared 
(Lenzen \cite{Le91}; Walsh et al. \cite{Wa99}). In addition,
millimeter wave continuum emission (Reipurth et al. \cite{Re96}) and molecular
line detection of CS have been reported 
(Zinchenko et al. \cite{Zi95}).
 A recent multi transitional study of IRAS 09002-4732
has led to the detection of bipolar outflow emission in the CS line
with distinct red and blue shifted peaks displaced symmetrically 
relative to the central IR object (Lapinov et al. \cite{La98}).
However, no bipolar emission in the CO
line has been seen by them.

The observations of a large region (area $>$ 500 sq. arc min) centered
at IRAS 09002-4732 are presented here.
This Galactic \ion{H}{ii} region / molecular cloud complex
has been studied 
with the aim of understanding the energetics,
physical sizes, spatial distribution of interstellar dust
and its temperature. 
  The next section describes the observations and the 
results alongwith the discussions are presented in section 3.

\section{Observations}
\subsection{Balloon-borne observations}

The Galactic star forming region associated with IRAS 09002-4732
was mapped using a new 12 channel two band far infrared (FIR)
photometer system at the Cassegrain focus of the  TIFR 100 cm (f/8) balloon 
borne telescope.  
Details of the 100 cm telescope system and the
observational procedure have been described 
elsewhere (Ghosh et al. \cite{Gh88};
Mookerjea et al. \cite{Mo99}). 
This 12 channel photometer was flown for the first
time on 18 November, 1993
from the TIFR Balloon Facility, Hyderabad (Latitude
= 17\fdg 47 N , Longitude = 78\fdg 57 E).
A pair of six element (2 $\times$ 3 close packed configuration) 
composite Silicon 
bolometer arrays was used as FIR detectors and were
cooled to 0.3 K using a closed cycle
$^{3}$He refrigerator (Verma et al. \cite{Ve93}). 
The same region of the sky was viewed simultaneously
in two FIR bands with near identical fields of view of
1\farcm 6 per bolometer, thus instantaneously
covering an area of 6\farcm 0 $\times$ 3\farcm 4 in each band.
A series of filters and a CsI beam splitter 
define the two FIR wavebands.
The spectral transmission of the 
total system in each FIR band was determined 
in the laboratory using a Michelson interferometer and
a Golay cell as the reference detector. 
The resulting transmission curves for the two bands
are shown in Fig. 1.
Since the configuration of the cryogenically cooled filters
which define the FIR pass bands, have been upgraded from time to time,
these bands differ from those in Mookerjea et al. (\cite{Mo99}).

%%%%%%%%%%%%%%%%%%%%%%% Fig1 begins
\begin {figure}
\resizebox{\hsize}{!}{\includegraphics{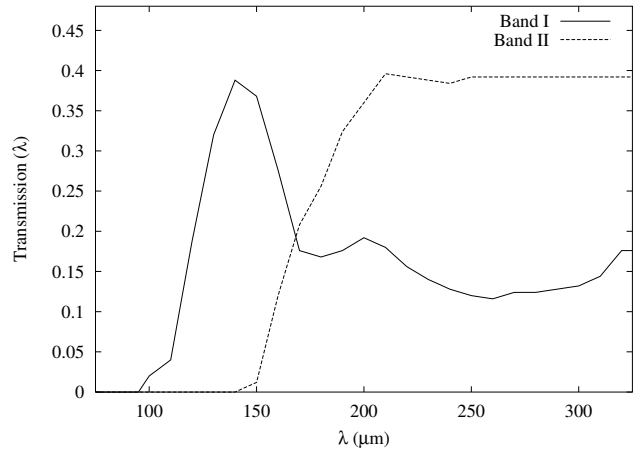}}
\caption{
The relative responses of the two far infrared bands of the photometer
used onboard the 1-meter balloon borne TIFR telescope.}
\end {figure}
%%%%%%%%%%%%%%%%%%%%%%% Fig1 Ends

The sky was chopped along the cross-elevation axis at 10 Hz 
with a throw of 4\farcm 2. 
The relative responses of individual bolometers, absolute flux 
calibration of the photometer as well as of the instrumental point 
spread function (PSF) were determined from raster scans across
Jupiter. The PSF also includes the effect of the sky chopping.
The calculated effective wavelengths 
and measured sensitivities for the two FIR bands
for various relevant grey body spectra
of different temperatures are presented in Table 1.
Hereafter, the two bands will be referred to as 148 and 209 $\mu$m
bands corresponding to the $\lambda_{eff}$ for a 30 K
source with $\lambda^{-2}$ emissivity.
The region of the sky $\approx$ 27\arcmin ~$\times$ 20\arcmin
~, centered on IRAS 09002-4732 was mapped twice.
The mapped region also included the IRAS Point Source Catalog 
(hereafter PSC) sources 
IRAS 08583-4719, 08589-4714 and 09014-4736.

%-------------------- Table 1---begins 
\begin{table*}
\caption{Characteristics of the two band FIR photometer.}
\vskip 0.5cm
\begin{tabular}{|c|c|c|c|c|c|c|c|c|}
\hline
T  &
\multicolumn{4}{c|}{Band-I} &
\multicolumn{4}{c|}{Band-II} \\
\cline{2-9}
      &
\multicolumn{2}{c|}{$\epsilon_{\lambda} \propto \lambda^{-1}$} &
\multicolumn{2}{c|}{$\epsilon_{\lambda} \propto \lambda^{-2}$} &
\multicolumn{2}{c|}{$\epsilon_{\lambda} \propto \lambda^{-1}$} &
\multicolumn{2}{c|}{$\epsilon_{\lambda} \propto \lambda^{-2}$} \\
%--------
\cline{2-9}

 &
$\lambda_{eff}$ 
& Relative & 
$\lambda_{eff}$ 
 & Relative & 
$\lambda_{eff}$ 
 & Relative &
$\lambda_{eff}$ 
 & Relative \\

(K)& 
$(\mu$m) & Sensitivity & 
$(\mu$m) & Sensitivity & 
$(\mu$m) & Sensitivity &
$(\mu$m) & Sensitivity  \\
\hline

 10 &    223 &  0.46 &  207 & 0.58 &   250 & 0.85 &    242 & 0.95  \\

 15 &    186 &  0.79  & 172 & 0.90 &   233 & 0.97 &    225 & 1.04  \\

 20 &    169 &  0.93  & 159 & 0.99 &   225 & 1.00 &    217 & 1.04  \\

 25 &    161 &  0.98  & 152 & 1.01 &   220 & 1.01 &    212 & 1.02  \\

 30 &    156 &  1.00  & 148 & 1.00 &   217 & 1.00 &    209 & 1.00  \\

 35 &    153 &  1.00  & 146 & 0.99 &   215 & 1.00 &    208 & 0.99  \\

 40 &    151 &  1.00  & 144 & 0.98 &   214 & 1.00 &    206 & 0.98  \\

\hline
\end{tabular}
\end{table*}
%-------------------- Table 1---ends

We carried out phase sensitive detection of individual    
bolometer signal using an on-board signal processing system.
The samples from different bolometers were combined during the
off-line data processing pipeline which reconstructs the
telescope aspect. This uses the relative
responses of the bolometers and a 
geometric model of their locations in the telescope focal plane.
The resulting final chopped FIR signals were gridded into 
a two dimensional sky matrix (elevation $\times$ cross elevation) with
a cell size of 0\farcm 3$\times$ 0\farcm 3. 
The observed signal map was 
deconvolved using an indigenously developed procedure based on the
Maximum Entropy Method (MEM) similar to that of Gull
\& Daniell (\cite{Gu78}) (see Ghosh et al. \cite{Gh88}, for details). 
The determination of the absolute aspect of the telescope 
was improved by using a focal plane optical photometer which 
detects stars (in an offset field) while the telescope scans the
programme source.
An absolute positional accuracy of $\sim$ 0\farcm 5 has been achieved
in the FIR maps using this method.
From the deconvolved maps of Jupiter,
the FWHM sizes  are found to be 
1\farcm 0$\times$1\farcm 4  and 1\farcm 0$\times$1\farcm 3  
in the 148 and 209 $\mu$m  bands respectively. 

\subsection{Data from IRAS and ISO missions}

  The data from the IRAS survey in the four bands 
(12, 25, 60 and 100 $\mu$m)
for the region of the sky mapped by us 
were HIRES processed (Aumann et al. \cite{Au90}) at the
Infrared Processing and Analysis Center (IPAC\footnote{IPAC is
funded by NASA as part of the part of the IRAS extended mission
program under contract to JPL.}, Caltech).
In the present study, these maps have been used for extracting 
sources and quantifying interband positional associations and flux densities.
They have also been used for generating maps of dust colour temperature
and optical depth.
Since both IRAS 09002-4732 and 09014-4736 appear in the
IRAS Low Resolution Spectrometer catalog, their spectra in the 
wavelength range 8--22 $\mu$m have also been used to construct 
the respective SEDs.
Similarly between 45--195 $\mu$m, the data from the Long Wavelength Spectrometer
onboard ISO have also  been used.

%%%%%%%%%%%%%%%%%%%%%%% Fig2 begins
\begin {figure*}
\resizebox{\hsize}{!}{\includegraphics{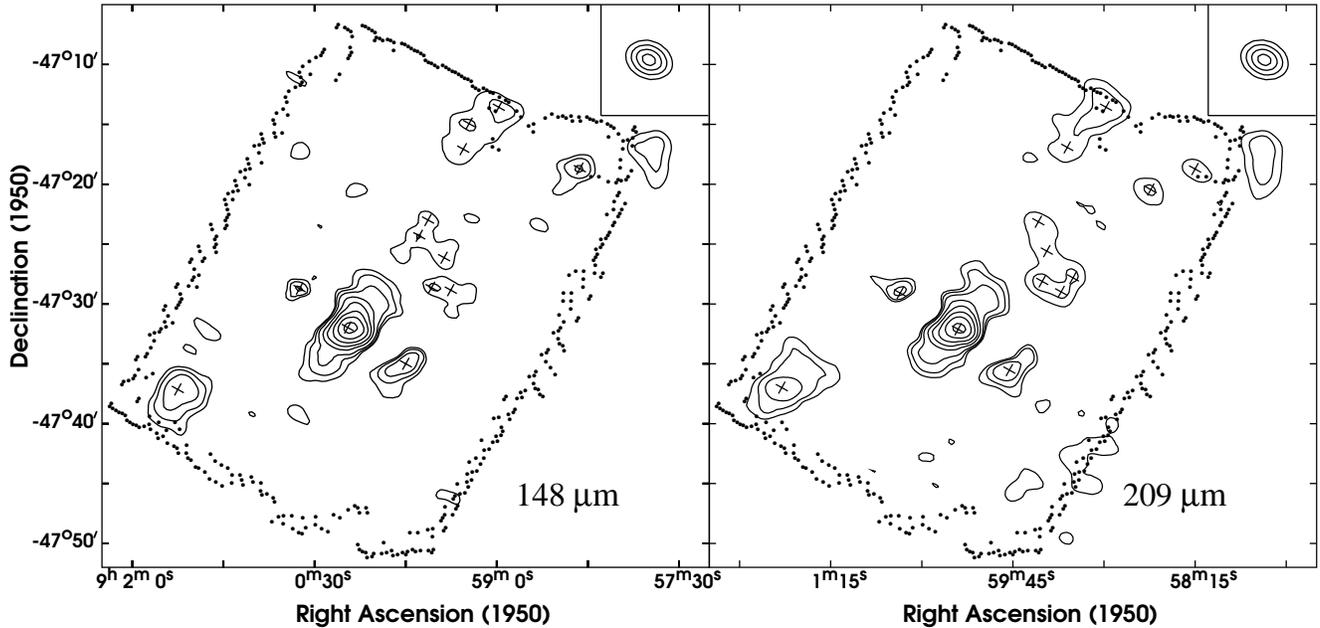}}
\caption{
The intensity maps for the region around IRAS 09002-4732 in TIFR
bands --
{\bf (a)} at 148 $\mu$m with peak = 3079 Jy/sq. arcmin,
{\bf (b)} at 209 $\mu$m  with peak = 1611 Jy/sq. arcmin.
The isophot contour
levels in both (a) and (b) are 80, 40, 20, 10, 5, 2, 1,
\& .5 \% of the
respective peaks. 
The crosses mark the positions of sources detected in respective
bands (see Sect. 3.1).
The insets show deconvolved images of Jupiter in the respective bands.
The contours are 60, 20, 5 \& 1\% of respective peaks, aligned to the
instrumental axes for meaningful comparison.  
}
\end {figure*}
%%%%%%%%%%%%%%%%%%%%%%% Fig2 Ends

\section{Results and Discussion}

\subsection{148 $\mu$m and 209 $\mu$m bands}

The MEM deconvolved TIFR maps 
at 148 and 209 $\mu$m 
of the Galactic star forming region
associated with IRAS 09002-4732 
are presented in Fig. 2 (a \& b). 
The insets show the 
deconvolved intensity contours for the point-like source (Jupiter),
to quantify the achieved angular resolution.
The dotted lines in these figures mark the boundary of the region 
covered by the telescope bore-sight. However the actual observations extend
beyond this boundary due to the sky chopping and the effective size of
the PSF array.
As a result, the intensity maps generated by the MEM deconvolution 
extend further by 3\farcm 4 along in-scan and 6\arcmin
~along cross-scan directions.
The limits for reliable extraction of discrete sources and structural 
information in the form of isophot contours have been 
established by using the data from two independent sets of raster scans.
The dynamic range has been found to be $\approx$ 200 for both bands.
Several regions of extended emission are seen in both the 148 and the
209 $\mu$m maps. A total of 15 sources have been detected, 
details of which are listed in Table 2 and positions marked on the maps.
The four most prominent peaks are associated with 
IRAS PSC sources, viz.,
08583-4719, 08589-4714, 09002-4732 and 09014-4736. 
The listed flux densities have been obtained by 
integrating over a circle of 3\arcmin ~diameter.

 The positions of the global peaks in both the 148 and 209 $\mu$m maps 
are identical to that of IRAS 09002-4732.
 The direction of the major extended emission seen from 
IRAS 09002-4732 in both the TIFR maps 
[size at 10 \% (50 \%) of peak : 
3\farcm 8 $\times$ 2\farcm 9 (1\farcm 8 $\times$ 1\farcm 6)
at both these bands]
matches quite well with mm-wave molecular line maps.
The CS (2--1) map of Zinchenko et al. (\cite{Zi95}) shows a FWHM size
of $\approx$ 2\arcmin $\times$ 1\farcm 5 with the major
axis at almost identical position angle as the FIR extension.
Higher angular resolution map by Lapinov et al. (\cite{La98}) in
the CS (7--6) line resolves the central region into two peaks 
with different LSR velocities nearly symmetrically displaced
relative to the FIR peak. The line joining these two peaks
also has a similar position angle.

\subsection{IRAS bands}

The HIRES processed maps in all the four
IRAS bands for the region of present study are shown 
in Fig. 3 (a,b,c \& d).
The dynamic range of these HIRES
maps is larger than that of the TIFR maps, and 
the isophots have been shown down to 0.1\%
of respective peak intensities.
Due to this larger dynamic range, the diffuse emission 
is more prominently visible in these HIRES maps.

 The angular resolutions achieved in these maps 
(in-scan $\times$ cross-scan) 
are 0\farcm 52 $\times$0\farcm 63, 
0\farcm 58$\times$0\farcm 70,
1\farcm 15$\times$1\farcm 35  and 1\farcm 87$\times$
1\farcm 97 at 12, 25, 60 and 100 $\mu$m respectively.

%%%%%%%%%%%%%%%%%%%%%%% Fig3 begins
\begin{figure*}
\resizebox{\hsize}{!}{\includegraphics{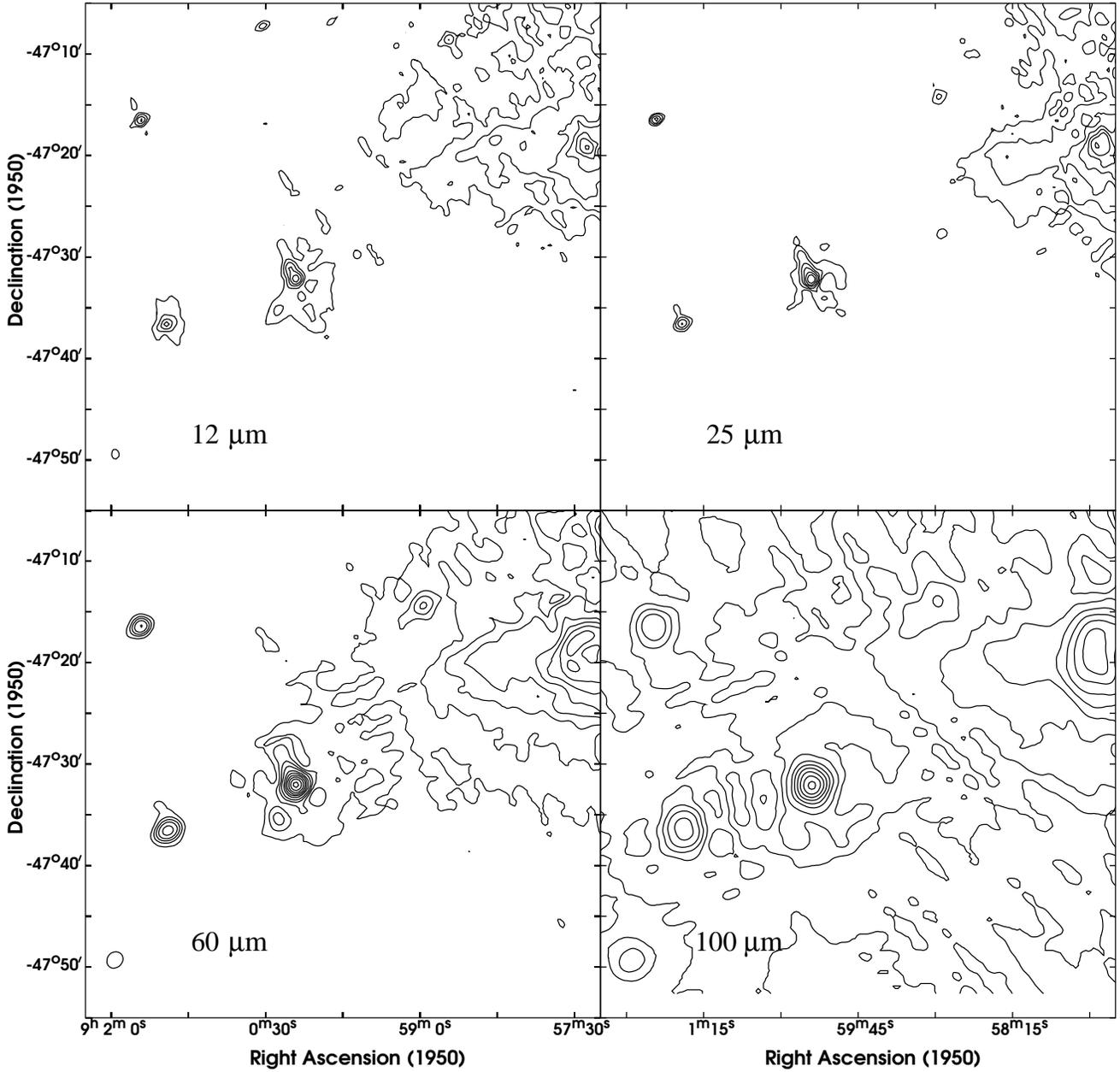}}
\caption{
The HIRES processed IRAS maps for a similar region as shown in Fig. 2,
in the four bands --
{\bf (a)} at 12 $\mu$m with peak = 374 Jy/sq. arcmin,
{\bf (b)} at 25 $\mu$m  with peak = 5470 Jy/sq. arcmin.
{\bf (c)} at 60 $\mu$m  with peak = 12300 Jy/sq. arcmin.
{\bf (d)} at 100 $\mu$m  with peak = 5830 Jy/sq. arcmin.
The isophot contour levels in (a) \& (b) are 
90, 30, 10, 3, 1, .3, \& .1 \% 
and in (c) \& (d) are 
80, 40, 20, 10, 5, 2.5, 1, .5, .25 \& .1 \% 
of the respective peaks.
}
\end{figure*}
%%%%%%%%%%%%%%%%%%%%%%% Fig3 Ends

 A new algorithm has been used to extract 
sources in the HIRES processed IRAS survey maps. The scheme works
on the local maxima in the intensity maps as candidates and models
the local background (encompassing the source of an assumed size, which is
a variable : 1--4\arcmin ~) and subtracts the same. 
The flux densities (3\arcmin ~diameter) for these HIRES sources are also 
presented in Table 2.

\subsection{Discrete sources }

 Out of the total 15 FIR sources that have been detected from the 
TIFR maps, 12 are common to both the 209 $\mu$m and 148 $\mu$m bands
(Table 2). 
All these 15 sources are found to be associated with
sources extracted from the HIRES maps in at least one band
(13 in at least 2 HIRES bands).
The positional match demanded is within 1\farcm 5 ~for the 
purpose of association.
Four of these are associated with IRAS PSC sources whose SEDs are 
studied in greater detail (Sect. 3.6).
The IRAS PSC flux densities for these four sources are also 
listed in Table 2.
The dust temperatures in the FIR, T${}_{FIR}$,
have been computed from the flux densities in the TIFR bands,
assuming an emissivity
law of $\epsilon_{\lambda} \propto {\lambda}^{-2}$. These are also
listed in Table 2.

 An attempt has been made to determine $\gamma$,
the slope of the IMF 
($dN/d(log M) \propto M^{\gamma}$), 
for the region around IRAS 09002-4732, 
under reasonable assumptions. We assume all the 15 sources 
to be nearly at the same distance based on the finding of WB89
(3 out of the 4 IRAS PSC sources in this region for which CO
emission has been found, are at distances within $\pm$ 10 \%).
If each source represents one embedded ZAMS star, and their
luminosities are translated to stellar masses using Tout et al. (\cite{To96}),
then the local slope of the IMF, $\gamma$ has been found to be
 $-1.25^{+0.75}_{-0.65}$ for the mass range 4--16 $M_{\odot}$.
The effect of incompleteness near the sensitivity limit has been 
avoided by appropriate choice of the mass bins based on the 
source number distribution.
This slope is similar to those found for Galactic OB associations
(Massey et al. \cite{Ma95}) and clusters (Phelps \& Janes \cite{Ph93}).

%---------------- Table 2 begins --------------------
\begin{table*}
\caption{Position and flux density details of the detected sources.}
\vskip 0.5cm
\begin{tabular}{ccccccccccc}
\hline
\#& RA & Dec & IRAS PSC&\multicolumn{6}{c}{Flux Density (Jy)} & T${}_{FIR}^{b}$ \\
\cline{5-10}
& (1950)    &(1950)  &  associations & \multicolumn{2}{c}{TIFR} & 
\multicolumn{4}{c}{IRAS$^a$} &\\
& h ~~m ~~s & o ~~\arcmin ~~\arcsec &&    209 $\mu$m & 148 $\mu$m&   
100 $\mu$m& 60 $\mu$m& 25 $\mu$m& 12 $\mu$m & (K) \\
\hline
S1& 8 58 15.0 & -47 18 44& 08583-4719 & 54 &209 & 1242 & 1309 & 233 
& 44 &$>$45\\
&"&"&"&&&          $<$1692$^c$ & $<$469$^c$ &    34.1$^c$ & 6.8$^c$ &\\
S2& 8 58 36.9 & -47 20 27&                 & 70 &--  &  767 &   -- 
&  36 & 25.3 &\\
S3& 8 58 59.0 & -47 13 35& 08589-4714 & 142& 195&  680 &  
365 & 30.1&  15.1 &21\\
&"&"&"&&&          551$^c$ & 200$^c$ &    13.5$^c$ & 2.2$^c$ &\\
S4& 8 59 13.7 & -47 14 57&                 &  --& 179&  459 &   -- 
&  -- & 10.7 &\\
S5& 8 59 15.3 & -47 27 50&                 &  90& 118&   -- &   
44 &  -- &  3.7 &20 \\
S6& 8 59 16.6 & -47 17 05&                 &  83& 155&   -- &  
110 &  -- & 10.9 & 27\\
S7& 8 59 22.8 & -47 28 51&                 & 102& 144&  197 &   
55 &  -- &  5.3 & 21\\
S8& 8 59 26.1 & -47 26 08&                 &  72& 121&   -- &   
43 &  -- &  -- & 24\\
S9& 8 59 30.5 & -47 28 07&                 & 102& 111&  187 &   
54 &  -- & 3.4 & 18\\
S10& 8 59 32.4 & -47 23 08&                 &  65& 102&   53 &   
43 &  -- &  -- & 23\\
S11& 8 59 38.3 & -47 24 18&                 &  --& 134&   53 &   
43 &  -- &  -- & \\
S12& 8 59 45.0 & -47 35 00&                 & 200& 384&  111 &   
46 &  17 &  6.3 & 28\\
S13& 9  0 12.1 & -47 32 07& 09002-4732 & 4753& 9127& 14057 & 13611& 
2577& 203 & 28\\
&"&"&"&&&        14707$^{c}$ & 11877$^{c}$ &   1962$^{c}$ & 121$^{c}$ &\\
S14& 9  0 40.7 & -47 29 06&                 &  66& 111  & --  &  
27 &  --  & -- & 25\\
S15& 9  1 39.1 & -47 37 00& 09014-4736 & 282& 607  & 902 & 
795 &  100 &  52 & 33 \\
&"&"&"&&&          957$^{c}$ & 621$^{c}$ &    79.6$^{c}$ & 30.0$^{c}$ &\\
\hline
\end{tabular}
\vskip 0.3cm
${}^a$ From HIRES processed maps unless specified otherwise. The 
flux densities are integrated over a circle of 3\farcm 0 diameter.

${}^b$ Determined using the flux densities in TIFR bands and assuming a
grey body spectrum with emissivity $\epsilon_{\lambda} \propto {\lambda}^{-2}$.

${}^c$ From IRAS Point Source Catalog.

\end{table*}
%---------------- Table 2 Ends --------------------

\subsection{Diffuse emission}

Using the T(148/209) temperature distribution for the entire region 
mapped (Sect. 3.5), and the intensity maps in the same region, the
total far-infrared luminosity has been estimated, assuming 
the dust emissivity law to be $\epsilon_{\lambda} \propto \lambda^{-2}$.
For the regions of the intensity maps with S/N $<$ 3 in either band,
a fixed value of 10 K has been assumed for the dust temperature.
The contribution of diffuse emission in this complex
is found to be $\sim$2.8$\times$10$^{4}$ $L_{\odot}$, which is 
35 \% of the total emission.
This does not include contribution from any diffuse (structureless
at $<$ 10\arcmin) component which may be associated with
the Vela Molecular Ridge clouds along the line of sight.
This fraction is in agreement with the values found for young Galactic 
star forming regions, viz., 25 \% for Orion A region (Mookerjea
et al. \cite{Mo00b}); 35 \% for W 31 complex (Ghosh et al. \cite{Gh89}).

\subsection{Distribution of dust temperature and optical depth}

Although the HIRES processed IRAS maps have a much higher dynamic range,
the angular resolution of the TIFR maps is superior to the IRAS
maps at 60 and 100 $\mu$m, due to the smaller and circular
beams employed.
Taking advantage of the nearly identical beams of
the TIFR bands at 148 and 209 $\mu$m and the
simultaneity of observations, we have generated reliable 
maps of dust temperature
and optical depth (at 200 $\mu$m, $\tau_{200}$; translated from 209 $\mu$m),
assuming a dust emissivity of
$\epsilon_{\lambda}\propto\lambda^{-2}$ following the same procedure
as described in Mookerjea et al. (\cite{Mo99}, \cite{Mo00b}). 
The dust temperatures are calculated based on the assumption
that the dust is optically thin which is reasonable at these wavelengths.
Considering the
uncertainties in the flux densities measured by us and the wavelength
regime probed by us, the temperature computed has an uncertainty of $\sim$
$\pm$2K, between 15 K and 35 K which gradually increases to 
$\pm$5K at 50 K.

%%%%%%%%%%%%%%%%%%%%%%% Fig4 begins
\begin{figure}
\resizebox{\hsize}{!}{\includegraphics{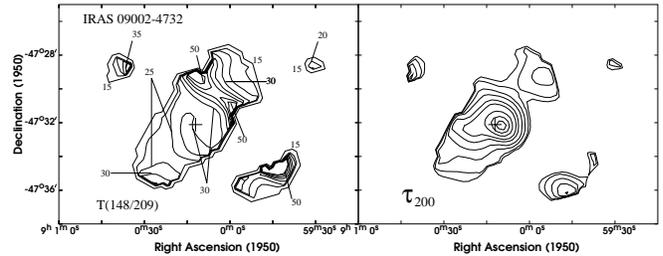}}
\caption{
The distribution of dust temperature T(148/209), and optical depth at 
200 $\mu$m, $\tau_{200}$, for the region around
IRAS 09002-4732 (A) assuming a dust emissivity law of 
$\epsilon_{\lambda} \propto \lambda^{-2}$.
The isotherms correspond to 15 K to 50 K in steps of 5 K.
Temperature values for a few contours are displayed to
avoid ambiguity.
The $\tau_{200}$ contours represent 90, 70, 50, 30, 20, 10, 5, 2 \& 1 \% 
of the peak value of 0.051.
The cross marks the position of IRAS 09002-4732.
}
\end{figure}
%%%%%%%%%%%%%%%%%%%%%%% Fig4 Ends

Two regions with relatively stronger FIR emission
have been selected out of the entire
mapped area for presenting the T(148/209) and $\tau_{200}$ distribution. 
These two regions cover the target source IRAS 09002-4732 
(hereafter, A) and the neighbouring 
source IRAS 09014-4736 (B).
The spatial distribution of T(148/209) as isotherm contours and contours
of constant $\tau_{200}$, for 
the regions A and B are presented in Figs. 4 and 5 respectively.

The HIRES maps at 60 and 100 $\mu$m 
trace the distribution of somewhat warmer dust (30--50 K).
We have used these to generate maps of the colour temperature, $T(60/100)$,
and optical depth ($\tau_{100}$) for 
the region A (around IRAS 09002-4732).
  The intensity maps at 60 and 100 $\mu$m were spatially averaged before 
computing $T(60/100)$ and $\tau_{100}$ in a manner similar to 
that described by Ghosh et al. (\cite{Gh93}) for  
an emissivity law of ${\epsilon}_{\lambda} \propto \lambda^{-1}$.
This emissivity law is more appropriate for the 60--100 $\mu$m wavelength
range.
The isotherms of T(60/100) and contours of constant $\tau_{100}$ are
presented in Fig. 6 (a \& b). 

The position of the intensity peaks in both the TIFR as well as all
four IRAS bands for the region A are identical to that of IRAS PSC
coordinates of IRAS 09002-4732, which has been marked on Figs. 4 and 6.
It may be noted here that the positions of the associated radio continuum 
source at 843 MHz (Whiteoak \cite{Wh92}) and H$_2$O maser source 
are within 15\arcsec ~of IRAS 09002-4732 (Lapinov et al. \cite{La98}).

The T(148/209) temperature distribution is 
very asymmetrical around IRAS 09002-4732.
Hotter dust and a steeper temperature 
gradient are seen towards the north-west (NW) of IRAS 09002-4732
and a relatively smoother temperature distribution in
the range of 20 -- 30 K is observed towards the south-east. 
The band of hotter dust is situated $\sim$ 2\farcm 2 to
the NW of the peak position of the FIR intensity.
The dust temperature drops from a value of $\sim$ 50 K to 20 K 
within a distance of $\approx$ 0.8 pc.
Similar large variations in dust temperature has been seen
for the regions associated with IRAS 00494+5617 and IRAS 05327-0457
(Mookerjea et al. \cite{Mo00a}).

The T(60/100) map shows one major peak, coinciding with the
position of IRAS 09002-4732 and the temperature drops (to 25 K)
on all sides almost symmetrically except for the region to the north. 
The values of T(60/100) would come closer to T(148/209), if the dust emissivity
exponent is chosen to be -2 (see Fig. 6 caption).
The lack of complex structure in this T(60/100) map vis-a-vis T(148/209)
map may be due to the poorer angular resolution of the IRAS maps.

The distribution of optical depth ($\tau_{200}$) around IRAS 09002-4732,
somewhat resembles 
the intensity distribution : the peak position matches with that of
$I_{148}$ or $I_{209}$. The peak $\tau_{200}$ value is 0.051. There is
a secondary peak in this map shifted towards the NW (beyond the
band of higher temperature), positionally matching with relatively cooler
dust [T(148/209) $\sim$ 20 -- 25 K].

%%%%%%%%%%%%%%%%%%%%%%% Fig5 begins
\begin{figure}
\resizebox{\hsize}{!}{\includegraphics{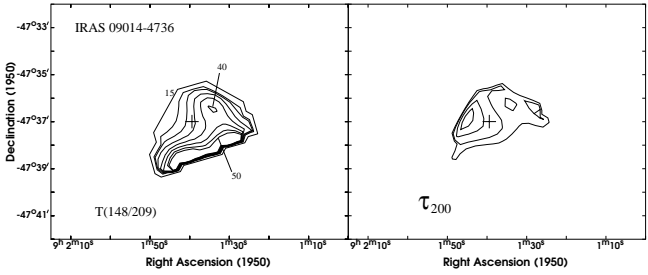}}
\caption{
Same as in Fig. 4, but for the region around IRAS 09014-4736 (B).
The isotherms refer to the same temperatures as in Fig. 4.
The $\tau_{200}$ contours represent 10, 5, 2 \& 1 \% of the global
peak value of 0.051.
The cross marks the position of the TIFR source (S15) associated
with IRAS 09014-4736.
}
\end{figure}
%%%%%%%%%%%%%%%%%%%%%%% Fig5 Ends

The $\tau_{100}$  distribution is similar to that of $\tau_{200}$
only near the peak of the intensity maps.
This may not be surprising, since the IRAS and TIFR maps are
probing dust column densities at different temperatures. 
The highest value of $\tau_{100}$ occurs
$\sim$ 2\arcmin ~south-west of IRAS 09002-4732.
The value of $\tau_{100}$ near the peak of the intensity maps, is 
0.023. 

The T(148/209) temperature distribution in IRAS 09014-4736 also
shows asymmetric structures (Fig. 5a). 
A ridge of hot dust runs approximately east-west 
$\sim$ 1\farcm 5 away (south) from the peak of FIR intensity distribution.
In addition, a local hotspot is seen due north-west (1\farcm 2) from 
the intensity peak.
The $\tau_{200}$ distribution shows one major 
peak positionally corresponding to $\sim$ 20 K dust, and an isolated
minor peak at the  position of the hotspot.

%%%%%%%%%%%%%%%%%%%%%%% Fig6 begins
\begin{figure}
\resizebox{\hsize}{!}{\includegraphics{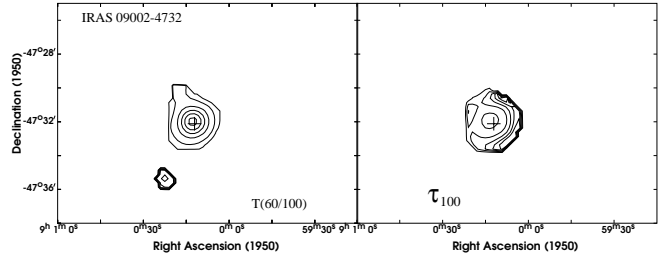}}
\caption{
The {\bf (a)} dust temperature T(60/100) 
and {\bf (b)} optical depth at 100 $\mu$m ($\tau_{100}$) distribution 
from the HIRES 
maps for the region around IRAS 09002-4732 (A)
assuming a dust emissivity law of 
$\epsilon_{\lambda} \propto \lambda^{-1}$.
The isotherms displayed in (a) correspond to the temperatures 70, 60, 50, 40, 30
\& 25 K. The innermost contour is for the hottest dust (70 K).
(The isotherms would represent 50, 45, 39, 33, 26 \& 22 K if the 
emissivity index was -2).
The contours in the optical depth map (b) represent 
50, 40, 30, 20 \&  10 \%  of the global
peak value of 0.11.
The cross marks the position of IRAS 09002-4732.
}
\end{figure}
%%%%%%%%%%%%%%%%%%%%%%% Fig6 Ends

\subsection{Radiative transfer modelling}

With the aim of extracting important physical parameters,
an attempt has been made to construct radiative
transfer models for the four IRAS PSC sources IRAS 09002-4732, 09014-4736, 
08583-4719 and 08589-4714, which have been covered in our FIR maps. 
The self consistent scheme developed by
Mookerjea \& Ghosh (\cite{MG99}) has been used for this purpose which is 
briefly described below.
This scheme models spherically symmetric cloud 
with centrally embedded young stellar object/(s), immersed in an average 
interstellar radiation field.
Although the gas-dust coupling is not considered explicitly,
the role of the gas in the energetics of the cloud has been included
in a self consistent manner.
Two commonly used types of interstellar dust have been explored
(Draine \& Lee \cite{Dr84}, hereafter DL; Mezger et al. \cite{Me82}, 
hereafter MMP), with a power law size distribution given by Mathis et al.
(\cite{Ma77}). For each type of dust, 
the relative proportion
of graphite and astronomical silicate grains is selectable.
The observational constraints include the SED due to the dust,
angular sizes at different wavelengths and the radio continuum emission.

The following parameters are explored in order
to get an acceptable fit to all the data :
(i) the nature of the embedded source, which could either be a 
 single ZAMS star or a cluster of ZAMS stars consistent with the
 Salpeter Initial Mass Function;
(ii) radial density distribution law (only three power laws
have been explored, viz., $n(r) \propto r^{\alpha}$, with 
$\alpha$ = 0, $-1$ or $-2$);
(iii) the relative abundances of the two constituent grain types;
(iv) total radial optical depth due to the dust (inclusive of all
constituents) at a selected wavelength ($\tau_{100}$ at 100 $\mu$m); 
(v) the gas to dust ratio by mass (the predicted radio 
continuum emission is sensitive to this);
(vi) geometric details of the dust cloud (e.g. cavity size, outer
size of the cloud).

 The distances to all the four IRAS PSC sources in the present study,
have been taken to be 1.4 kpc.

\subsubsection{IRAS 09002-4732}

 The observed SED of IRAS 09002-4732 has been constructed using TIFR,
IRAS-HIRES, IRAS-LRS and ISO-LWS data (Fig. 7a).
In addition, 1.3 mm continuum measurement of 
Reipurth et al. (\cite{Re96}) and near infrared data from Lenzen (\cite{Le91}) have
been used.
The total radio continuum emission has been found to be 3.8 Jy at 5 GHz
(Caswell \& Haynes \cite{Ca87}) from this source in a wide beam 
(4\farcm 4).
Whiteoak (\cite{Wh92}) found the flux density at 843 MHz to be 2.1 Jy.
Recent high angular resolution (synthesized beam 
$\approx$ 2\arcsec ~) 
measurement for the central region 
by Walsh et al. (\cite{Wa98}) at 8.64 GHz, gives a peak intensity of
0.5 Jy/beam. 
The total luminosity and the rate of Lyman continuum photon generation 
has been estimated to be 6.9 $\times 10^{4} L_{\odot}$ and
7.1 $\times$ $10^{47}$ s${}^{-1}$ respectively by 
Simpson \& Rubin (\cite{Si90}).
By integrating the observed SED, we estimate the total luminosity to be
1 $\times 10^{5} L_{\odot}$.
In a study of a sample of ultracompact \ion{H}{ii} regions which includes
IRAS 09002-4732, Walsh et al. (\cite{Wa99}) 
have fitted the infrared part of the SED by representing it 
as a two-component black body spectrum and also by a radiative transfer model.
Our self consistent modelling scheme includes angular sizes at 
different wavelengths, radio continuum data as well as dust
composition and grain size distribution.

  The best radiative transfer model for IRAS 09002-4732 has been
identified by varying the parameters and 
comparing the following predictions with observations :
(i) the SED originating from the dust, (ii) the angular sizes at different
wavebands, and (iii) the radio continuum emission from the gas component.
The predicted angular sizes have been 
convolved with the relevant telescope beams, prior to comparison
with observed FWHM sizes.
The embedded source is found to be a single ZAMS O7 star with
$N_{Lyc}$ = 3.6 $\times 10^{48}$ s$^{-1}$ and $T_{eff}$ = 38,500 K.
The preferred type of dust is DL and the density distribution is uniform. The 
gas to dust ratio is 100 : 1 by mass.
The other parameters of the best fit model for IRAS 09002-4732 are 
presented in Table 3.

%----------------Table 3 begins ----------------------------------------------
\begin{table*}
\caption{Parameters for the best fit radiative transfer models.}
\vskip 0.5cm
\begin{tabular}{ccccccccc}
\hline
Source & $\alpha$& $R_{max}$& $R_{min}$& $r_{H II}$& $\tau_{100}$&Luminosity
& Dust composition$^{a}$ &$M_{dust}$\\
&&(pc)&(pc)&(pc)&&($10^{3} L_{\odot}$)& Silicate:Graphite & $(M_{\odot}$)\\
\hline
IRAS 08583-4719& 0.0&0.3& $5.0\times 10^{-5}$& $2.9\times 10^{-3}$ & 0.02& 4.0 & 81 : 19 &0.50\\
IRAS 08589-4714& 0.0&0.2& $8.0\times 10^{-5}$& $1.2\times 10^{-3}$& 0.05& 2.4 & 81 : 19 &0.55\\
IRAS 09002-4732&  0.0&1.0&$3.0\times 10^{-2}$& $5.5\times 10^{-2}$& 0.05& 100 & 72 : 28 &14\\
IRAS 09014-4736& -1.0&0.8& $8.0\times 10^{-5}$& $2.9\times 10^{-5}$ & 0.03& 6.5 & 81 : 19 &0.80\\
\hline
\end{tabular}
\vskip 0.3cm
$^{a}$Dust is of DL type for all the four sources.
\end{table*}
%----------------Table 3 ends----------------------------------------------

 The predicted SED from this model is compared with the observations
in Fig. 7 (a). The model fits the data quite well.
The predicted total radio continuum emissions (1.5 Jy at 843 MHz; 
3.7 Jy at 5 GHz \& 3.5 Jy at 8.6 GHz) are also in reasonable agreement
with observations. 
For our model, 
the \ion{H}{ii} region extends into the dust cloud 
beyond the dust-free central cavity ($r_{H II} > R_{min}$), and its
size is of similar order as 
inferred from the 8.6 GHz map by Walsh et al. (\cite{Wa98}).
A comparison of the predicted angular sizes at mid to far infrared 
wavelengths with the measurements is presented in Table 4.
The predicted angular sizes increase as a function of the wavelength
and the values are in general agreement within errors.

 The IRAS LRS catalog has classified IRAS 09002-4732 to be of type 81
based on the strengths of emission of the unidentified 11.3 $\mu$m feature
and the [Ne II] line at 12.8 $\mu$m. Our inference of an embedded O7 
star from the above modelling is consistent with the detection of
the [Ne II] line. However, the detection of 11.3 $\mu$m feature,
generally attributed to be arising from single photon absorption
processes of PAH molecules, indicates importance of non-equilibrium
heating processes,
which is beyond the scope of the present study.

The total mass of our model cloud is 1400 $M_{\odot}$ (gas to dust ratio = 100)
which can be compared 
with inferences from molecular line observations.
The virial mass estimate from the CO data of WB89 and using 
the relation of Evans (\cite{Ev99}; see equation 8), we get the
cloud mass to be 1710 $M_{\odot}$ which is reasonably close to
the mass from our model considering the uncertainties.
On the other hand, the mass estimate from CS line measurements is
607 $M_{\odot}$ (Zinchenko et al. \cite{Zi95}).
The $(L/M)$ ratio of the total IR luminosity to cloud mass 
from the model, the virial mass based on CO data, and the 
CS core mass, are 
71, 58 and 165  $L_{\odot}/M_{\odot}$ respectively, for IRAS 09002-4732.

%-------------Table 4 begins ---------------------------
\begin{table}
\caption{Comparison of predicted sizes from the best fit model 
for IRAS 09002-4732
with observed  FWHMs.}
\vskip 0.5cm
\begin{tabular}{|c|c|c|}
\hline
 Wavelength & Model & Observations \\
 ($\mu$m) & (~\arcmin ~) & (~\arcmin ~) \\
\hline
 12  & 0.59 & 0.54 \\
 25  & 0.61 & 0.62 \\
 60  & 1.07 & 0.89 \\
 100 & 1.96 & 1.33 \\
 148 & 2.4 & 1.8 \\
 209 & 2.5 & 1.8 \\
\hline
\end{tabular}
\end{table}
%-------------Table 4 ends ---------------------------

\subsubsection{IRAS 09014-4736}

 The FIR emission from IRAS 09014-4736 at 148 and 209 $\mu$m is quite 
extended as evident from Fig. 2  and it is 
listed in the IRAS Small Scale Structure
Catalog (SSSC; X0901-476) due to its extended emission in 25 and 60 $\mu$m
bands.
It is gratifying to note that the flux densities
listed in the SSSC (109 Jy at 25 $\mu$m and 794 Jy at 60 $\mu$m) 
match extremely well with flux densities (100 and 795 Jy respectively;
see S15 in Table 2) determined using our source extraction
scheme from the HIRES processed maps (Sect. 3.2).
The angular sizes estimated in SSSC are : 1\arcmin ~at 25 $\mu$m
and 1\farcm 6 at 60 $\mu$m. The FWHM sizes (major $\times$ minor) in the 
148 and 209 $\mu$m bands are 3\farcm 5 $\times$ 2\farcm 7 and
3\farcm 8 $\times$ 2\farcm 5 respectively. The size increases
with wavelength as expected.

 The observed SED for IRAS 09014-4736 has been compiled using the TIFR,
IRAS-HIRES and IRAS-LRS data (Fig. 7b). No radio continuum data for 
this source are available in the literature. However, this region has
been covered by the 843 MHz survey with 43\arcsec ~angular resolution
(Whiteoak \cite{Wh92}). The quoted completeness limit 
of 50 mJy by Whiteoak (\cite{Wh92})
has been used here as an upper limit.
Modelling the SED of
IRAS 09014-4736 leads to a better fit for an $r^{-1}$ density 
distribution law than for an $r^0$ law.
The embedded source is found to be a ZAMS star 
of B1 type ($T_{eff}$ = 22,600 K).
The preferred dust is of DL type with 
the radial optical depth, $\tau_{100}$ = 0.03.
The other parameters corresponding to the
best fit model for IRAS 09014-4736 are listed in Table 3.
The spectral fit of this model to the observed SED is displayed
in Fig. 7 (b) which is reasonably good.
There is hardly any detectable radio continuum emission
as expected for a B1 type star.

Near and mid-IR spectrophotometry of IRAS 09014-4736 
shows recombination line emission of hydrogen (Persson \& Campbell \cite{Pe87};
Beck et al. \cite{Be91}; Porter et al. \cite{Po98}).
The LRS spectrum of this source has been classified as type 80 
based on the detection of 11.3 $\mu$m feature and absence of
any fine structure line from ionized heavier elements, typical
of low excitation \ion{H}{ii} region.
Detection of several additional PAH features (Cohen et al. \cite{Co89};
de Muizon et al. \cite{Mu90}; Zavagno et al.
\cite{Za92}) further supports that IRAS 09014-4736 has an embedded YSO.
Although IRAS 09014-4736 is classified as an unidentified source in 
the IRAS PSC, it is situated within $\approx$ 20\arcsec ~of
the small optical nebula, BBW 225 (size $<$ 2\arcmin ~),
identified by Brand et al. (\cite{Br86}).
The embedded YSO (B1) inferred from our modelling is consistent 
with all of the above. However, the actual geometry of the source
has to be far from spherically symmetric, in order to reconcile with the
size of the optical nebula and the near infrared data. Similar conclusions
have been arrived at by Beck et al. (\cite{Be91}) for IRAS 09014-4736.

The only available molecular line observation from this source is 
the CO emission (WB89).
We estimate 
the molecular gas density and the virial mass 
for the cloud associated with IRAS 09014-4736 to be
5.2 $\times$ 10$^3$ cm$^{-3}$ and 395 $M_{\odot}$ respectively.
The mass of the model cloud is 80 $M_{\odot}$ for a gas to dust mass 
ratio of 100. The latter is not constrained well due to the 
lack of radio continuum emission.
The $(L/M)$ ratio corresponding to the model and the virial 
mass of this cloud are
81 and 16 $L_{\odot}/M_{\odot}$ respectively.

%%%%%%%%%%%%%%%%%%%%%%% Fig7 begins
\begin{figure*}
\resizebox{\hsize}{!}{\includegraphics{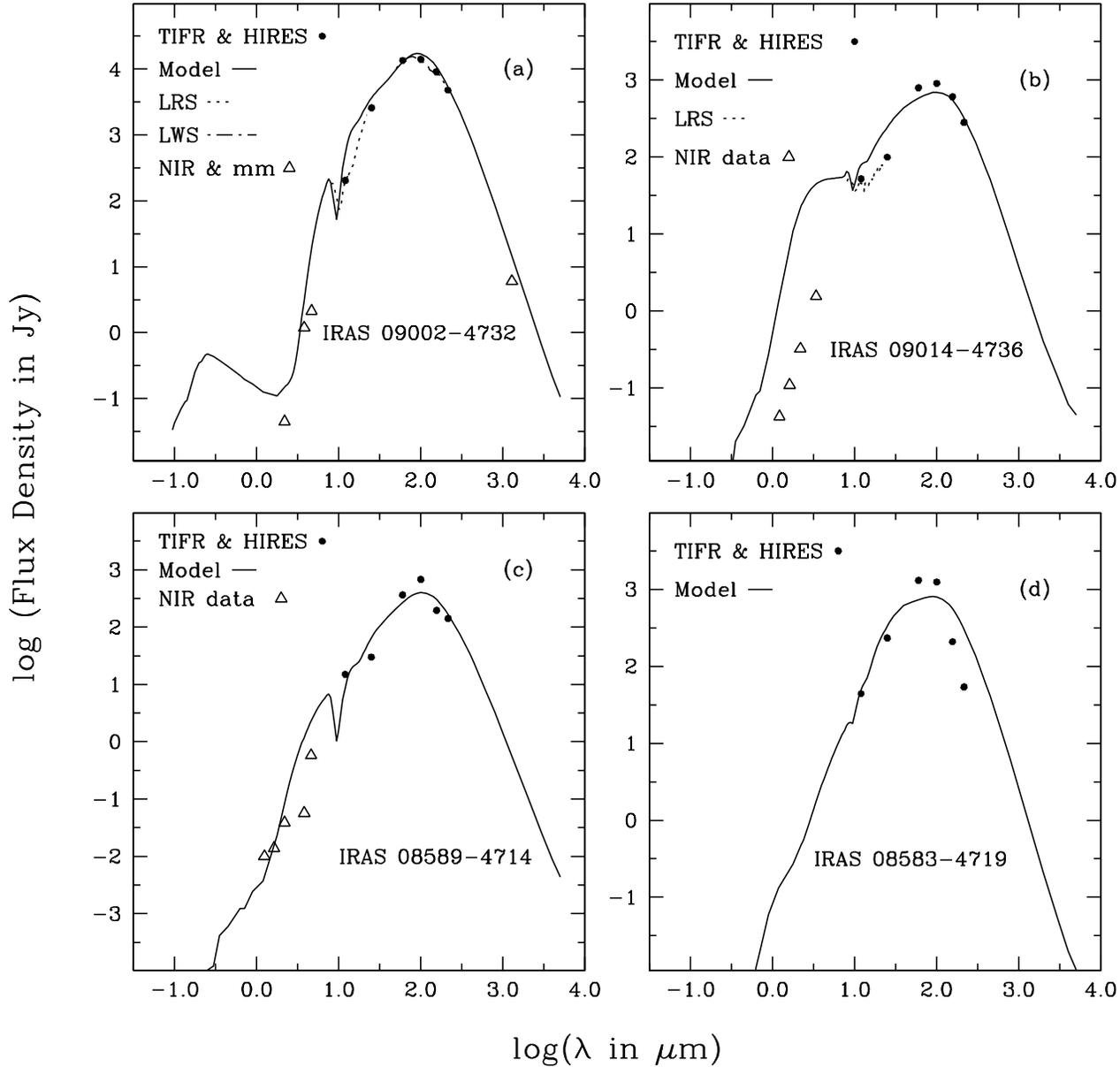}}
\caption{
Comparison of the predicted spectral energy distributions 
(SEDs) from the
best fit models of {\bf (a)} IRAS 09002-4732, {\bf (b)} IRAS 09014-4736
{\bf (c)} IRAS 08583-4719 \& {\bf (d)} IRAS 08589-4714
with observations. 
The filled circles represent the TIFR and IRAS-HIRES data.
The dashed lines are IRAS-LRS spectra (for IRAS 09002-4732
and IRAS 09014-4736) and the dash dot line is the ISO-LWS
spectrum of IRAS 09002-4732. The triangles denote 1.3 mm data
from Reipurth et al. (\cite{Re96}) for IRAS 09002-4732 and near-IR
data from Lenzen (\cite{Le91}) for IRAS 09002-4732, from Persson \&
Campbell (\cite{Pe87}) for IRAS 09014-4736, from Lorenzetti et al. (\cite{Lo93}) 
for IRAS 08589-4714.
The solid line denotes the prediction of the best fit model.
See text and Table 3 for details of the model parameters.
}
\end{figure*}
%%%%%%%%%%%%%%%%%%%%%%% Fig7 Ends

\subsubsection{IRAS 08589-4714}

 In addition to the measurements presented here, the near-IR
data of Lorenzetti et al. (\cite{Lo93}) have been used to construct
the SED of IRAS 08589-4714 (Fig. 7c).
The parameters of the best fit radiative transfer model are presented in
Table 3. A constant density cloud with DL type dust grains is implied.
The total radial $\tau_{100}$ = 0.05 and the model cloud mass is
 55 $M_{\odot}$. 
Since the energizing stellar source is detectable in the near-IR, this
is an example of very evolved stage of star formation, where only a small
fraction of the original interstellar matter is left around the star at
present.

  The source IRAS 08589-4714 has been 
detected in CO and CS surveys by Wouterloot \&
Brand (\cite{Wo89}) and 
Bronfman et al. (\cite{Br96}) respectively. 
Whereas our model gives a value for the $L/M$ ratio to be 
$\approx$ 44 $L_{\odot}/M_{\odot}$,
the virial mass of the
cloud estimated from the CO data of WB89 indicates a value
of 3.7 $L_{\odot}/M_{\odot}$.

\subsubsection{IRAS 08583-4719}

   IRAS 08583-4719 is contained in the catalog of bright rimmed clouds
which are candidates for star formation by radiation driven implosion
(Sugitani \& Ogura \cite{Su94}). Unfortunately no additional observational
data pertaining to a SED are available other than those presented here. 
The physical size estimate has been taken from 
Sugitani \& Ogura (\cite{Su94}).
 Once again a constant density dust (DL) cloud gives the best fit to the
SED (Fig. 7d). However, the fit to the observations even for this best model 
is not very satisfactory. 
The parameters for this model are listed in Table 3.
The mass of the model cloud is 50 $M_{\odot}$ leading to
a $(L/M)$ ratio of 
80 $L_{\odot}/M_{\odot}$.

\subsubsection{Inferences from the modelling}

 The DL type of dust is favoured for all the four sources
modelled here. It may be noted that the emissivity index 
in the 100 -- 200 $\mu$m range
 for DL type dust (for both silicate and graphite), is 
$\approx$ -2, consistent with our choice for generation of T(148/209) 
maps (Sect. 3.5).
 Three of these four sources suggest a constant density
dust envelopes around the YSOs. This conclusion seems to be very
common for several Galactic star forming regions : e.g. for IRAS 18314-0720,
18355-0532 and 18316-0602 (Mookerjea \& Ghosh, \cite{MG99}); IRAS 00338+6312,
04004-5114 and RAFGL 5111 (Mookerjea et al. \cite{Mo99}); IRAS 00494+5617
and 05327-0457 (Mookerjea et al. \cite{Mo00a}).
The following comments can be made by
comparing the parameters 
for the four sources modelled here :
The dust cloud corresponding to IRAS 09002-4732 has 
the largest size of the central cavity, which contains the most
luminous embedded source.
Except IRAS 09014-4736, for all other three sources, the \ion{H}{ii} regions
extend well into the respective dust clouds. This emphasizes the role
of dust grains on the emission of recombination lines and radio continuum.
The radial optical depth due to the dust ranges between 0.02 to 0.05
which is slightly on the lower side compared to IRAS 00494+5617
(Mookerjea et al. \cite{Mo00a}), 00338+6312 (Mookerjea et al. \cite{Mo99}).
The interstellar dust composition is found to be nearly identical in 
all the four sources modelled, viz., Silicate dominated.
The ratio between the
graphite and the silicate grains, in general vary drastically
among the Galactic star forming regions (Mookerjea et al. \cite{Mo99}, \cite{Mo00a}).
Hence, nearly identical dust composition found here further supports
the hypothesis that the entire complex around IRAS 09002-4732 mapped
by us is physically associated.

The $L/M$ ratio for the four sources studied here, as found 
from our radiative transfer modelling, falls in a narrow range 
of 44 -- 81 $L_{\odot}/M_{\odot}$.
For three of these for which CO measurements are available, 
this ratio based on virial masses, fall in somewhat broader range 
(3.7 -- 58 $L_{\odot}/M_{\odot}$).
These can be compared with the results from other studies :
(i) A study similar to the present one, 
for the W 31 complex has lead to a value of $(L/M)$
in the range of 1.6 -- 3.6 
$L_{\odot}/M_{\odot}$ (Ghosh et al. \cite{Gh89});
(ii) the average value for a large sample of clouds using virial mass
based on CO measurements, is
$\sim$ 4 $L_{\odot}/M_{\odot}$ with a spread exceeding
a factor of 100 (Mooney \& Solomon \cite{Mo88}, Evans \cite{Ev91});
(iii) for denser cores studied in CS emission (Plume et al. \cite{Pl97})
give $\sim$ 190 $L_{\odot}/M_{\odot}$ with a spread of a factor of 15.
Considering the uncertainties arising due to the fact that
the different tracers are
sampling different components of the ISM (at possibly different 
spatial scales), the star forming complex associated with
IRAS 09002-4732 can be termed as typical of the Galaxy.

\section{Summary}

  The Galactic star forming region associated with the
IRAS source 09002-4732 and its neighbouring region ($\approx$ 0.15 sq. deg.)
has been mapped in two far infrared bands at 148 and 209 $\mu$m
with an angular resolution of $\sim$ 1\arcmin .
A total of 15 sources have been detected including 4 IRAS PSC
sources. Several of these are well resolved in both these bands.
Diffuse emission contributes 35 \% of the total FIR emission 
at 209 $\mu$m.
Taking advantage of simultaneous observations using identical beams 
in these two bands, reliable maps of dust colour temperature [T(148/209)] and
optical depth ($\tau_{200}$) have been generated which show
many structures. 
Under reasonable assumptions, the IMF slope for this region
has been found to be $-1.25^{+0.75}_{-0.65}$ for the mass 
range of 4--16 $M_{\odot}$.

The HIRES processed IRAS survey data for the same region 
at 12, 25, 60 and 100 $\mu$m have also been used for comparison
as well as extracting discrete sources positionally matching
with those detected at 148 and 209 $\mu$m.
The spatial distribution of T(60/100) and $\tau_{100}$ have also been
generated for the region locally around IRAS  09002-4732
for comparison. 

  Self consistent  radiative transfer calculations has been carried out
for IRAS 09002-4732, 09014-4736, 08583-4719 and 08589-4714 through
spherical dust-gas clouds.
The observed infrared sub-mm SED, angular sizes and the radio continuum data
have been used to constrain the parameters of the models.
Whereas for IRAS 09014-4736 an $r^{-1}$  
radial density distribution law is preferred, all the other three
sources favour an $r^{0}$ law.
The geometrical details of the dust-gas clouds, 
the dust compositions and optical depth, etc.
have been quantified by these models. 
The luminosity per unit mass for these sources lie in a rather
narrow range of 44--81 $L_{\odot}/M_{\odot}$.
  
\begin{acknowledgements}

We thank 
S.L. D'Costa, M.V. Naik, S.V. Gollapudi, D.M. Patkar, 
M.B. Naik and G.S. Meshram for their support for the
experiment. The members of TIFR Balloon Facility (Balloon group and
Control \& Instrumentation group), Hyderabad, are thanked for their roles
in conducting the balloon flights. IPAC is thanked for providing HIRES
processed IRAS data. We thank ISO Data Centre for making the
ISO-LWS data public.
 We thank the referee Dr. J. G. A. Wouterloot whose
suggestions have improved the scientific content of this paper.

\end{acknowledgements}

%%%%%%%%%%%%%%%%%%%%%%%%%%%% End of Main Text %%%%%%%%%%%%%%%%%%%%%%%%%%%%%%

%%%%%%%%%%%%%%%%%%%%%%%%%%%% Bibliography begins %%%%%%%%%%%%%%%%%%%%%%%%%%%

%%%%%%%%%%%%%%%%%%%%%%%%%%%% Bibliography ends %%%%%%%%%%%%%%%%%%%%%%%%%%%%%

\end{document}